# Ferroelectrically Switchable Half-Quantized Hall Effect

M. U. Muzaffar,[§,†] Kai-Zhi Bai,[‡] Wei Qin,[¶] Guohua Cao,[§,†] Bo Fu,[#] Ping Cui,[§,†,*] Shun-Qing Shen,[‡,*] and Zhenyu Zhang[§,†,*]

[§]*International Center for Quantum Design of Functional Materials (ICQD), Hefei National Research Center for Physical Sciences at the Microscale, University of Science and Technology of China, Hefei, Anhui 230026, China*

[†]*Hefei National Laboratory, University of Science and Technology of China, Hefei, Anhui 230088, China*

[‡]*Department of Physics, The University of Hong Kong, China*

[¶]*Department of Physics, University of Science and Technology of China, Hefei, Anhui 230026, China*

[#]*School of Sciences, Great Bay University, Dongguan, Guangdong 523000, China*

*E-mail: cuipg@ustc.edu.cn; sshen@hku.hk; zhangzy@ustc.edu.cn

## ABSTRACT

Integrating ferroelectricity, antiferromagnetism, and topological quantum transport within a single material is rare, but crucial for developing next-generation quantum devices. Here, we propose a multiferroic heterostructure consisting of an antiferromagnetic $MnBi_2Te_4$ bilayer and an $Sb_2Te_3$ film is able to harbor the half-quantized Hall (HQH) effect with a ferroelectrically switchable Hall conductivity of $\pm e^2/2h$. We first show that, in the energetically stable configuration, the antiferromagnetic $MnBi_2Te_4$ bilayer opens a gap in the top surface bands of $Sb_2Te_3$ through proximity effect, while its bottom surface bands remain gapless; consequently, HQH conductivity of $e^2/2h$ can be sustained clockwise or counterclockwise depending on antiferromagnetic configuration of the $MnBi_2Te_4$. Remarkably, when applying interlayer sliding within the $MnBi_2Te_4$ bilayer, its electric polarization direction associated with parity-time reversal symmetry breaking is reversed, accompanied by a reversal of the HQH conductivity. The proposed approach offers a powerful route to control topological quantum transport in antiferromagnetic materials by ferroelectricity.

Over the past decades, there have been rapid developments in two important areas of condensed matter physics, namely, quantum anomalous Hall effects[1-5] and ferroelectricity[6-9], wherein symmetry breaking plays an integral part in the emergence of different quantum states [10]. The former, a fundamental transport phenomenon characterized by either integer[2, 3, 11-14] or half-integer[15-20] quantum Hall conductance in multiples of $e^2/h$ (where $e$ is the electron charge and $h$ is Planck's constant), arises from time-reversal symmetry ($\hat{T}$) breaking, while the latter results from inversion symmetry ($\hat{P}$) breaking, leading to spontaneous polarization[21]. Notably, unlike the integer quantum anomalous Hall effect in insulating phases, semi-magnetic topological insulators with a single gapless Dirac cone exhibit a half-integer quantized Hall conductivity, accompanied by a non-zero longitudinal resistance[16]. Recently, the broken $\hat{P}$ symmetry in ferroelectric topological insulators has been actively explored, as it combines switchable polarization with robust topological surface states [22-25], opening new avenues for designing innovative functional devices. The simultaneous breaking of both $\hat{T}$ and $\hat{P}$ symmetries may lead to even more exotic phenomena than breaking either $\hat{T}$ or $\hat{P}$ symmetry alone[26]. In particular, ferroelectrically switchable and topologically protected edge states in a quantum regime may pave the way for developing next-generation quantum devices with high fault tolerance[27].

Recently, $MnBi_2Te_4$ has attracted significant attention as an intrinsic magnetic topological insulator with antiferromagnetic interlayer coupling, where the spins of the magnetic elements are ferromagnetically aligned within the $ab$ plane[28-31]. Consequently, the total magnetization depends on the film thickness, with films containing odd numbers of layers exhibiting the integer quantum anomalous Hall state[32], while those containing even number of layers exhibiting zero net magnetization and no anomalous Hall conductivity[33]. Here, the existence or absence of the quantum anomalous Hall effect can again be simply formulated based on symmetry arguments. Such a symmetry approach has recently led to the discovery of an asymmetric Hall response in the even layers of $MnBi_2Te_4$, driven by $\hat{P}\hat{T}$ symmetry breaking through external electric fields[34] or by intrinsic ferroelectricity via layertronic sliding [35], countering the conventional wisdom that the anomalous Hall effect is tied to net magnetization. In particular, the demonstration of the electrical switching ability of the Hall conductance is not only a rare achievement due to conflicting symmetry requirements but also holds great promise for next-generation high-performance antiferromagnetic spintronic devices. However, to fully realize the application potentials, it is crucial to go beyond the existing paradigm that primarily focuses on insulating phases by exploring alternative routes to achieve nonvolatile switching of the quantized Hall conductivity.

In this Letter, by going beyond the existing recipe in ferromagnetic systems[16, 36], we propose a multiferroic $MnBi_2Te_4/Sb_2Te_3$ heterostructure with both antiferromagnetic and ferroelectric orders as an ideal platform for realizing the half-quantized Hall (HQH) conductivity with versatile tunability. We first show that, in the energetically stable configuration, the antiferromagnetic $MnBi_2Te_4$ bilayer opens a gap in the top surface bands of $Sb_2Te_3$ through proximity effect, while its bottom surface bands remain gapless; consequently, a HQH conductivity of $e^2/2h$ can be sustained clockwise or counterclockwise depending on the initial antiferromagnetic configuration of the $MnBi_2Te_4$ bilayer. More strikingly, when applying a relative interlayer sliding within the $MnBi_2Te_4$ bilayer, its electric polarization direction associated with the $\hat{P}\hat{T}$ symmetry breaking is reversed, accompanied by a reversal of the HQH conductivity as well. The proposed approach offers a powerful route to control topological quantum transport in antiferromagnetic materials by ferroelectricity, with appealing technological potentials.

First-principles calculations were performed using the projector-augmented wave method[37], implemented in the VASP code[38]. The Perdew-Burke-Ernzerhof approximation was used to describe the exchange-correlation functional[39]. To treat the localized $3d$ orbitals of Mn atoms in our proposed heterostructure, the (GGA)+$U$ method[40] was employed, with the $U_{eff}$ = 4.0 eV, as suggested in Ref.[41]. Anomalous Hall conductivity was calculated using the WannierTools package[42] in conjunction with the Wannier90[43]. More computational details, along with the relevant references[37-50], are given in the Supporting Information.

Tetradymite-type $MnBi_2Te_4$ forms a rhombohedral layered structure, similar to typical three-dimensional topological insulators (TIs) such as $Bi_2Te_3$, and belongs to the R3m space group (No. 166)[29]. Its basic building block is a septuple layer (SL), stacked in the sequence Te-Bi-Te-Mn-Te-Bi-Te (see Figure 1(a)). Within each SL, a long-range ferromagnetic is established, while adjacent SLs exhibit antiferromagnetic coupling, with a Néel temperature ($T_N$) of 24 K[51]. Due to the inherent $\hat{P}\hat{T}$ symmetry, a film with an even number of layers of $MnBi_2Te_4$ in an antiferromagnetic order is centrosymmetric. However, breaking space inversion symmetry could potentially lead to the emergence of ferroelectricity[7-9]. In doing so, we first consider anti-AA stacking of an isolated $MnBi_2Te_4$ bilayer by rotating the top layer 180° relative to the bottom layer to break the inversion symmetry of the system, as shown in Figure S1. Such a stacking has been demonstrated in various hexagonal 2D materials, including

MnBi$_2$Te$_4$[7, 8, 35, 50, 52]. Moreover, by further performing a lateral interlayer translation operation on the top layer of the anti-AA stacking with ± t$_∥$ (where t$_∥$ = [1/3, -1/3] in fractional coordinates, as shown in Figure S1), two additional stacking configurations, namely anti-AB and anti-AC, can be obtained. Such inversion symmetry-broken configurations are energetically stable, while the anti-AA configuration is metastable due to the proximity of Te atoms across the interface[35]. We then stack anti-AB and anti-AC configurations of MnBi$_2$Te$_4$ bilayers on thick Sb$_2$Te$_3$ to form semi-magnetic TIs, as shown in Figure 1(a). Our total energy calculations reveal that both the anti-AB and anti-AC based heterostructures are energetically nearly degenerate, exhibiting a camel-back feature in the total energy between the two states (see Figure 1(b)) that potentially signifies the emergence of ferroelectricity. The calculated binding energy for the anti-AB heterostructure is –27.01 meV/Å², comparable to that of typical vdW heterostructures[53]. Here, the binding energy is defined as $E_b = E_{total} − E_{MBT} − E_{ST}$, with $E_{total}$, $E_{MBT}$, and $E_{ST}$ being the total energies of the heterostructure, freestanding bilayer MnBi$_2$Te$_4$, and freestanding five quintuple layers (QLs) of Sb$_2$Te$_3$, respectively.

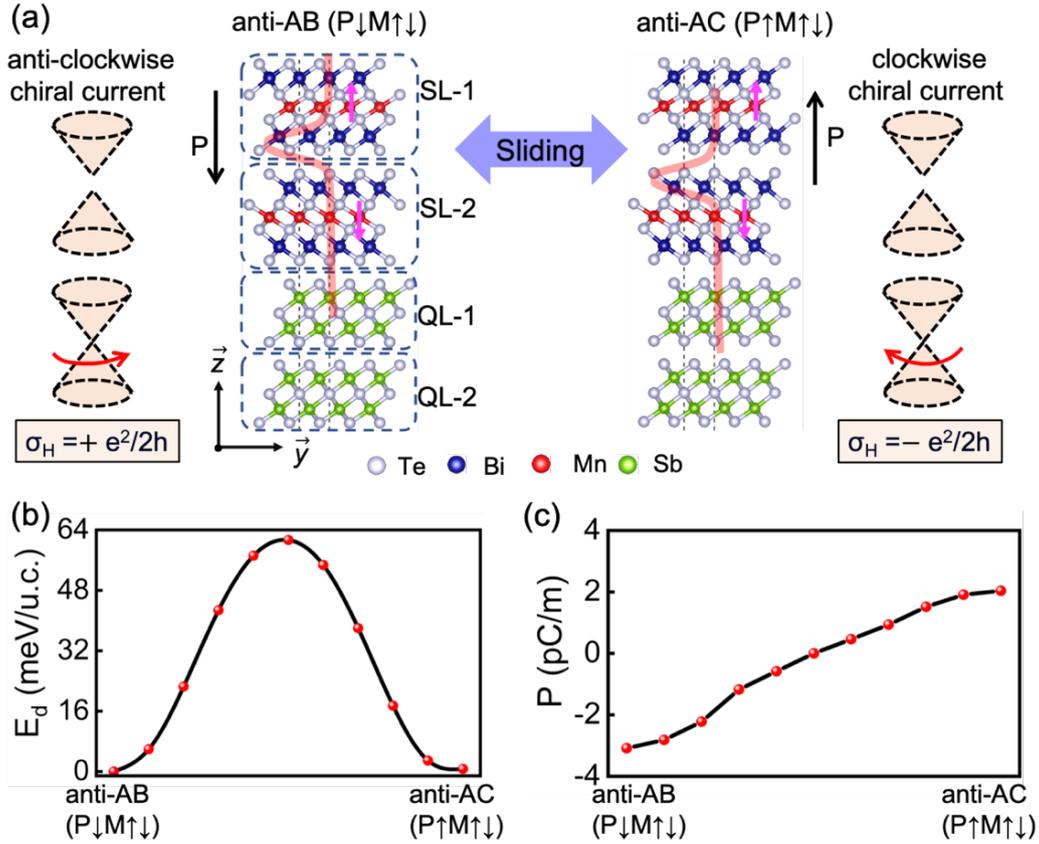

**Figure 1.** (a) Atomic structures of the two switchable ferroelectric states in the MnBi$_2$Te$_4$/Sb$_2$Te$_3$ heterostructure, along with the ferroelectric switching mechanism. In the two central columns, a schematic illustration is shown between the density distribution of the gapped surface states (as represented by the red lines within the crystals) and the tunability of the parity anomaly of the gapless Dirac cone at the bottom surface. The Mn $3d$ atoms are labeled in red, with opposite spin directions indicated by the magenta arrows and opposite polarizations shown by the black arrows. (b) Energy profile and (c) variation in the out-of-plane electric polarization during the ferroelectric switching of the MnBi$_2$Te$_4$/Sb$_2$Te$_3$ heterostructure.

Next, to quantify the spontaneous polarizations ($P$) in the two stacking heterostructures, we employ the Berry phase approach[54] and the ferroelectric switching pathway is determined via the nudged elastic band (NEB) method[55]. We find that the anti-AB and anti-AC heterostructures exhibit relatively high $P$ values of -3.08 pC/m and 2.03 pC/m, respectively, and the energy barrier between them is 61 meV per unit cell (u.c.) (see Figure 1(b)). This barrier is higher than that of isolated MnBi$_2$Te$_4$ bilayers (31 meV/u.c.) but still lower than or comparable to that of well-established 2D ferroelectric materials such as In$_2$Se$_3$ (66 meV/u.c.)[56]. Thus, MnBi$_2$Te$_4$/Sb$_2$Te$_3$ semi-magnetic TI exhibits robust ferroelectricity with switchable out-of-plane polarization, making it ideal for low-energy, non-volatile switching and memory devices.

Now we discuss the magnetic properties during ferroelectric polarization reversal due to magnetoelectric coupling in the MnBi$_2$Te$_4$/Sb$_2$Te$_3$ heterostructure. As shown in Figure 1(b,c) and Figure S2, all the stacking configurations of the MnBi$_2$Te$_4$/Sb$_2$Te$_3$ heterostructure along the kinetic pathway prefer out-of-plane antiferromagnetic ground state. Moreover, compared to the freestanding MnBi$_2$Te$_4$ bilayer (0.24 meV per Mn atom), the MnBi$_2$Te$_4$/Sb$_2$Te$_3$ heterostructure exhibits enhanced magnetic anisotropy, increasing to 0.33 meV and 0.39 meV per Mn atom for the anti-AB and anti-AC stackings, respectively. Typically, the magnetic ordering temperature of a material is closely related to its magnetic anisotropy, with a higher anisotropy leading to higher Curie or Néel temperature[57]. To confirm this conjecture, we have performed Monte Carlo simulations to estimate $T_N$ for the anti-AB (P↓M↑↓) configuration of the MnBi$_2$Te$_4$/Sb$_2$Te$_3$ heterostructure as a specific example (for details, see Section S1). Using the DFT-derived exchange interactions $J$ and magnetic anisotropy energy ($E_{MAE}$), the estimated $T_N$ for the anti-AB (P↓M↑↓) configuration is 38.7 K (see Figure S3), which is higher than that of an isolated MnBi$_2$Te$_4$ bilayer, given by 24 K[58].

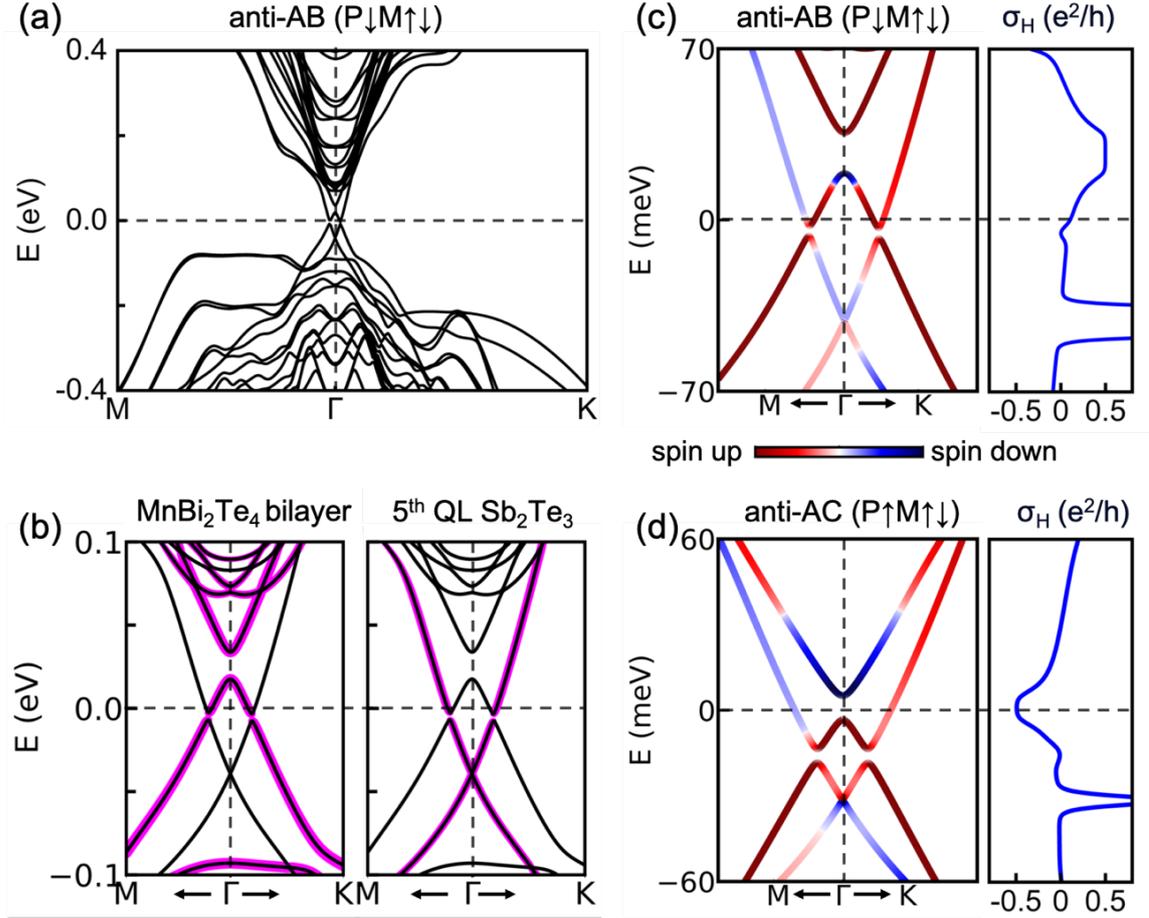

**Figure 2.** (a) Electronic structure and (b) zoomed-in views of the constituent band structures for the anti-AB (P↓M↑↓) configuration of the MnBi$_2$Te$_4$/Sb$_2$Te$_3$ heterostructure. (c,d) Electronic structures (left panels) and HQH conductivities (right panels) as functions of energy for the two ferroelectrically switchable states of the MnBi$_2$Te$_4$/Sb$_2$Te$_3$ heterostructure, with the projected expectation value of the spin operator $S_z$ also indicated by the spin bar.

Next, we turn our attention to the electronic structure and quantum transport properties of the two switchable states of the MnBi$_2$Te$_4$/Sb$_2$Te$_3$ heterostructure. The Sb$_2$Te$_3$ film possesses two gapless surface Dirac cones, and their mutual coupling is critically dependent on the film thickness[59]. To crosscheck that the 5-QL film of Sb$_2$Te$_3$ is thick enough, we project the wave functions of the topmost or bottommost QL onto the band structure, as shown in Figure S4. We see that the topologically nontrivial surface states cross the Fermi level, indicating sufficient film thickness to decouple the top and bottom surface states. Upon depositing the bilayer MnBi$_2$Te$_4$ on top of Sb$_2$Te$_3$, the exchange interaction between the surface electrons of the TI and magnetic elements in the MnBi$_2$Te$_4$ induces an energy gap ($\Delta$) in the top-surface Dirac cone of the TI while the bottom surface Dirac cone remain gapless, as shown in Figure 2(a,b). To distinguish between the two switchable states of anti-AB (P↓M↑↓) and anti-AC (P↑M↑↓), we calculate the spin-dependent band structures with spin-orbit coupling, as shown in Figure 2 (c,d). Remarkably, both configurations exhibit a spin-dependent magnetic gap around the Γ point, which varies upon reversing the direction of the electric polarization. Furthermore, as the proposed heterostructure also exhibits net antiferromagnetic order, when the directions of the magnetic moments are flipped while keeping the polarization direction fixed, the sign of the magnetic gap changes as well, as shown in Figure S5. Such a spin-dependent magnetic gap can serve as a control parameter for the anomalous Hall conductivity, as confirmed later. The calculated magnetic gap for the anti-AB (P↓M↑↓) and anti-AC (P↑M↑↓) configurations is 16.2 meV and 8.4 meV, respectively, and these values are tunable through strain engineering, as also demonstrated for the anti-AB (P↓M↑↓) configuration as an example (see Figure S6). Here it is noted that due to quantum confinement effect, a tiny gap appears below the Fermi level, and it decreases with the thicker Sb$_2$Te$_3$ layers; however, due to computational constraints, we restrict the analysis to up to 5-QLs.

The Hall conductivity at zero temperature can be calculated using the Kubo formula[60],

$$\sigma_H = -\frac{e^2}{h} \int_{BZ} \frac{dk_x dk_y}{2\pi} \Omega^z(k_x, k_y). \tag{1}$$

Here the Berry curvature is defined as

$$\Omega^z(k_x, k_y) = -2 \operatorname{Im} \sum_n^{occupied} \sum_{n'}^{empty} \frac{\langle \Psi_{nk} | \hbar v_x | \Psi_{n'k} \rangle \langle \Psi_{n'k} | \hbar v_y | \Psi_{nk} \rangle}{(\varepsilon_{nk} - \varepsilon_{n'k})^2}, \tag{2}$$

where $\Psi_{nk}$ is the Bloch wave function with eigenvalues $\varepsilon_{nk}$ and $v_x$ ($v_y$) is the velocity operator along the $x$ ($y$) direction. The calculated Hall conductivities for the two switchable ferroelectric states of the MnBi$_2$Te$_4$/Sb$_2$Te$_3$ heterostructure are plotted in Figure 2(c,d) where both configurations exhibit the HQH effect with a Hall conductance of magnitude $e^2/2h$ within the magnetic gap. In sharp contrast, an isolated MnBi$_2$Te$_4$ bilayer is a trivial insulator with zero Hall conductivity due to its inherent $\hat{P}\hat{T}$ symmetry (see Figure S9). The switchable HQH conductivity can be explained based on Berry curvature analyses, as presented in Figure 3. It can be seen that reversing the polarization direction from the anti-AC (P↑M↑↓) to the anti-AB (P↓M↑↓) state while keeping the magnetic moment directions fixed changes the sign of the Berry curvature around the Γ point, leading to a switchable Hall conductivity from $-e^2/2h$ to $+e^2/2h$. This controllable switching of the Berry curvature and the corresponding HQH conductivity through electric filed is not only experimentally feasible [34] but also highly desirable for device applications. These analyses show that switchable polarization and/or magnetization can be used as control parameters for realizing and manipulating the HQH effect.

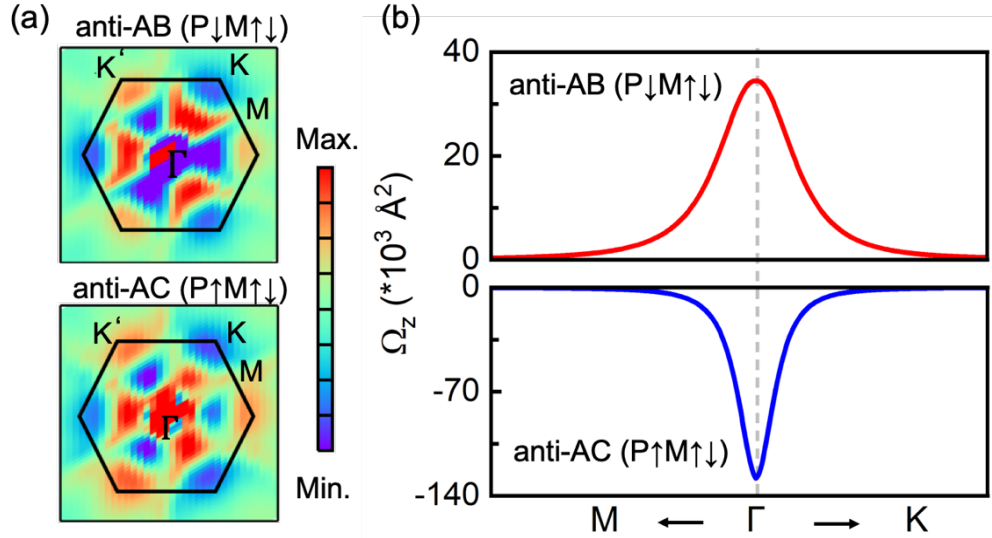

**Figure 3.** (a) Berry curvature distributions in the 2D Brillouin zone, and (b) along high-symmetry points, for the two ferroelectrically switchable states of the $MnBi_2Te_4/Sb_2Te_3$ heterostructure.

It is worth noting that, to realize HQH conductivity, an antiferromagnetic $MnBi_2Te_4$ bilayer serves as a prototypical example to open the energy gap in the top surface bands of $Sb_2Te_3$, while keeping the bottom surface bands gapless. This proposed approach for realizing HQH conductivity is conceptually generic and applicable to other related 2D semi-magnetic TIs composed of antiferromagnetic–nonmagnetic topological material combinations, provided the magnetization is strong enough to induce an energy gap in the top-surface bands of the TI while keeping the bottom-surface bands gapless (for details, see Section S3). As for $MnBi_2Te_4$, it typically exhibits intrinsic antisite defects that may influence the energy gap in the top-surface bands of TI,[61] and several effective methods[51, 62], such as slower cooling and additional thermal annealing, are commonly employed to drive the system towards thermodynamic equilibrium during growth and promote the formation of high-quality, uniform single crystals.

In Figure 4(a,b), we present the real-space charge density distributions of the upper-surface topological surface states at the Γ point for both the anti-AB (P↓M↑↓) and anti-AC (P↑M↑↓) configurations of the MnBi$_2$Te$_4$/Sb$_2$Te$_3$ heterostructure. It can be seen that a large peak in the real-space charge density distribution primarily appears in the bottom layer of MnBi$_2$Te$_4$ for the anti-AC (P↑M↑↓) configuration, namely, the MnBi$_2$Te$_4$ layer in proximity to the TI film. Upon reversing the polarization, the topological surface states of the TI in proximity with the MnBi$_2$Te$_4$ float up towards the top MnBi$_2$Te$_4$ layer. This floating behavior of the surface states leads to a strong interaction with the magnetism, resulting in a larger surface band gap observed in the anti-AB (P↓M↑↓) configuration compared to the anti-AC (P↑M↑↓).

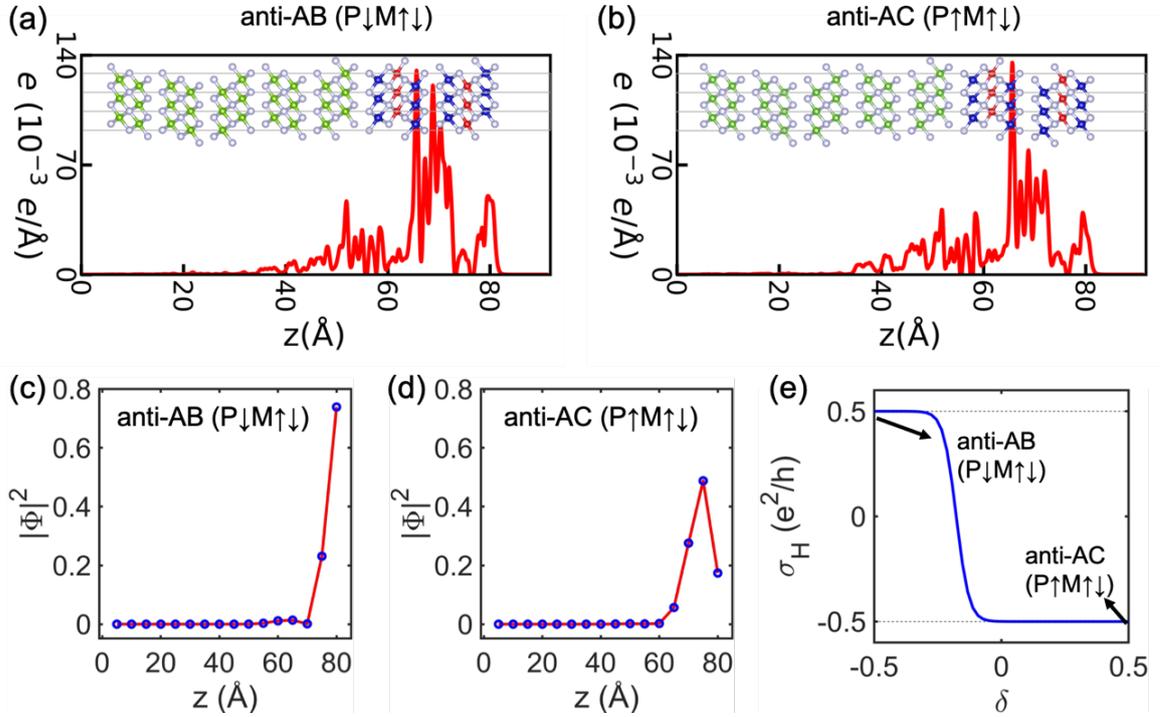

**Figure 4.** (a,b) Electron density distributions of the gapped topological surface states at the Γ point for both polarizations, with the crystal structures shown in the insets. (c,d) Density distributions of the gapped surface states for both ferroelectric states and (e) HQH conductivity as a function of inter-layer coupling strength (δ).

Going beyond first principles calculations, we employ a tight-binding lattice model to explore the impact of inter-layer coupling strength (δ) on the Hall conductivity in our proposed heterostructure (for details, see Section S2). In Figure 4(c-e), we present the calculated density distributions of the surface states and the Hall conductivity as a function of δ, revealing a quantum phase transition characterized by a switchable Hall conductivity of $\pm e^2/2h$. This quantum phase transition can be understood as the peak of the surface state density in the anti-AB (P↓M↑↓) configuration is localized at the first MnBi$_2$Te$_4$ layer; while in the anti-AC (P↑M↑↓) configuration, the peak shifts to the second MnBi$_2$Te$_4$ layer, consistent with our first-principles results. The sign of the magnetic gap is correlated with the parity anomaly of the gapless Dirac cone at the bottom surface of the TI film, which determines the sign of its half quantized Hall conductivity. These analyses suggest that the switching of ferroelectricity alters the dominance of density distribution within the MnBi$_2$Te$_4$ layers, leading to switchable Hall conductivity in the MnBi$_2$Te$_4$/Sb$_2$Te$_3$ semi-magnetic TI.

Before closing, we wish to emphasize the significance of this work from both the fundamental and practical aspects. First, our work explores the delicate interplay between the surface states of a TI and antiferromagnetism, uncovering a pathway to achieve HQH conductance that challenges the prevailing notion that antiferromagnetic systems with zero net magnetization yield no anomalous Hall effect. More importantly, the switching ability of the HQH conductance via ferroelectricity in the proposed semi-magnetic MnBi$_2$Te$_4$/Sb$_2$Te$_3$ heterostructure can significantly broaden the scope of practical device applications, especially compared to systems that rely on ferromagnetic phase[3, 15, 16, 32, 35]. Secondly, since observation temperatures of the quantum anomalous Hall effect have so far been limited to only a few kelvin[3, 32, 63, 64], significant efforts continue to be made to identify new quantum topological materials in the antiferromagnetic phase[65, 66]. Here, the proposed antiferromagnetic phase exhibits a large magnetic gap of 16.47 meV and a relatively higher Néel temperature of ~39 K suggesting that the observation temperature for the HQH effect in the present system can be much higher than what has been observed so far in the ferromagnetic phase (~1 K)[16]. From the practical aspect, the proposed MnBi$_2$Te$_4$/Sb$_2$Te$_3$ heterostructure is also physically realistic. Moreover, our detailed *ab initio* molecular dynamics simulations have shown that the predicted heterostructure is stable at least up to 200 K (see Figure S10), thereby offering a desirable temperature range for the fabrication of such systems and exploration of the emergent quantum transport properties.

In summary, following the structural and compositional compatibility between MnBi$_2$Te$_4$ and Sb$_2$Te$_3$, we have comprehensively demonstrated that the proposed multiferroic MnBi$_2$Te$_4$/Sb$_2$Te$_3$ heterostructures exhibit moderate sliding energy barriers and ferroelectrically switchable HQH conductivity at elevated temperatures. This central finding offers a highly viable route for designing low-power memory devices with antiferromagnetic spintronic applications at relatively high temperatures.

ACKNOWLEDGEMENTS

This work was supported by the National Natural Science Foundation of China (Grant Nos., 12374458, 12488101, 11974323, 12004368, 12474134, W2433002 and GG9990001216), the Innovation Program for Quantum Science and Technology (Grant No. 2021ZD0302800), the Strategic Priority Research Program of Chinese Academy of Sciences (Grant No. XDB0510200), the Anhui Provincial Key Research and Development Project (Grant No. 2023z04020008), the Research Grants Council, University Grants Committer, Hong Kong (Grant Nos. C7012-21G and 17301823), and Quantum Science Center of Guangdong-Hong Kong-Macao Greater Bay Area (Grant No. GDZX2301005).